\documentclass[runningheads]{llncs}

\usepackage{eccv}

\usepackage{eccvabbrv}

\usepackage{graphicx}
\usepackage{booktabs}
\usepackage[table,xcdraw]{xcolor}
\usepackage{array}
\usepackage{multirow}
\usepackage{adjustbox}
\usepackage[figuresright]{rotating} 
\usepackage[accsupp]{axessibility}  

\usepackage{hyperref}

\usepackage{orcidlink}

\begin{document}

\title{Real-World Multi-Modal and\\Longitudinal Lung Cancer Dataset} 

\titlerunning{Real-World Multi-Modal and\\Longitudinal Lung Cancer Dataset}

\author{Rita Cordeiro Mendes\inst{1}\orcidlink{0009-0005-0753-3269} \and
Maria Rita Fonseca Verdelho\inst{1}\orcidlink{0000-0002-1407-8322} \and
Carlos Santiago\inst{1}\orcidlink{0000-0002-4737-0020} \and 
Catarina Barata\inst{1}\orcidlink{0000-0002-2852-7723}}

\authorrunning{R. Cordeiro Mendes et al.}

\institute{Institute for Systems and Robotics, Instituto Superior Técnico, Portugal\\
\email{rita.cordeiro@tecnico.ulisboa.pt}
}

\maketitle
\begin{abstract}
  Multi-modal learning has demonstrated strong potential in medical applications by integrating heterogeneous data sources such as medical imaging, clinical records, and genomics to improve predictive performance and support clinical decision-making. However, advances in this area are often constrained by two key challenges: the limited availability of well-curated, ready-to-use datasets that accurately reflect real-world conditions, where medical data are frequently collected inconsistently and are often incomplete; and the inherent difficulty of integrating heterogeneous data modalities. In this work, we introduce a newly curated multi-center, multi-modal, and longitudinal dataset designed to support the evaluation of a wide range of learning pipelines under realistic conditions. The dataset comprises a total of 1,365 lung cancer patients and has three imaging modalities (whole-slide images, CT scans, and PET scans), structured clinical data, transcriptomic, and longitudinal follow-up and treatment information. For each imaging modality the dataset contains more than one instance. Moreover, the dataset exhibits substantial and non-uniform missingness across modalities, making it well-suited for studying robust multi-modal fusion strategies. We further provide both uni-modal and multi-modal benchmarks on the task of 12-month overall survival prediction, disease-specific survival, as well as longitudinal benchmark of hazard prediction under severe missing data. Our results show that, despite high levels of missingness, integrating complementary modalities consistently improves predictive performance over uni-modal approaches, highlighting the value of multi-modal fusion in realistic clinical settings.
  The dataset and benchmark code are available at \href{https://github.com/ritacmendes/MMIST-LUNG}{GitHub}.
\end{abstract}

\section{Introduction}
\label{sec:intro}
Healthcare is undergoing a profound transformation driven by digital technologies, the increasing availability of large-scale biomedical data, and advances in computational methods. In this context, multi-modal learning has emerged as a promising paradigm for leveraging the diverse sources of information routinely generated in modern clinical practice.

Traditional uni-modal systems, which rely exclusively on a single type of data, are often insufficient to capture the complexity of patient health \cite{breast_cancer_prog, scoping_review}. Each modality provides only a partial representation of the patient's condition and may fail to capture clinically relevant information that is available in other modalities.

These limitations underscore the need for models capable of integrating multiple heterogeneous data modalities, mirroring the multidisciplinary approach adopted in clinical practice, where specialists combine complementary expertise to develop a more comprehensive understanding of a patient's condition. Consequently, one of the central challenges in medical artificial intelligence is the effective integration of heterogeneous multi-modal information \cite{DELEN2005113, medical-prognosis}.

The ongoing digitalization of healthcare has facilitated the systematic collection and storage of diverse biomedical data, including medical images, genomic profiles, laboratory measurements, electronic health records, and clinical narratives. The growing availability of such multi-modal data offers unprecedented opportunities to develop more accurate, robust, and generalizable predictive models, thereby advancing the goals of precision medicine \cite{multimodal-fusion}. However, constructing high-quality multi-modal datasets remains a significant challenge. Beyond the sheer volume, dimensionality, and heterogeneity of the data, researchers must also address the considerable effort required to harmonize information originating from multiple independent sources. Furthermore, missing modalities are a pervasive issue, as not every patient undergoes the same diagnostic procedures or data acquisition protocols. As a result, real-world clinical datasets frequently exhibit incomplete multi-modal information, presenting a major obstacle to the development and deployment of reliable multi-modal learning systems.

Within the ecosystem of the National Cancer Institute (NCI), publicly available resources such as The Cancer Genome Atlas (TCGA), The Cancer Imaging Archive (TCIA), and the Clinical Proteomic Tumor Analysis Consortium (CPTAC) have become widely used due to their scale and extent, which covers imaging, omics, and clinical information. However, despite their value, these datasets are distributed across separate platforms and lack consistent curation and integration, which limits their accessibility and direct usability for multi-modal machine learning research.

In this work, we publicly release a curated multi-center, multi-modal, and longitudinal dataset of 1,365 lung cancer patients constructed from TCGA, TCIA, and CPTAC. The dataset integrates clinical variables, imaging data, genomic information, and longitudinal follow-up records, and reflects realistic clinical conditions, including extensive missingness across modalities. It also supports multiple observations per modality (e.g., repeated imaging studies and follow-up time points).

Moreover, we provide a set of uni-modal and multi-modal benchmarks for predicting 12-month overall patient survival, disease-specific survival as well as longitudinal hazard prediction in the lung cancer cohort.

Our contributions can be summarized as follows:
\begin{itemize}
    \item The release of a curated real-world multi-modal dataset comprising multiple instance of three imaging modalities (WSIs, CT, and PET), transcriptomic profiling, longitudinal follow-up and treatment information, and clinical data for 1,365 patients with substantial missingness, further detailed in Section \ref{sec:dataset}.
    \item A comprehensive set of benchmark methods for evaluating multi-modal approaches, while handling missing and incomplete modalities, including using a Transformer-based framework that naturally accommodates incomplete or fully missing modalities through attention masking \cite{aremultimodaltransformersrobust}.
\end{itemize} 

\section{Lung Dataset}
\label{sec:dataset}
Lung cancer remains one of the most prevalent and lethal malignancies \cite{lung_most_common}. Ongoing research efforts have led to the development of several large-scale public datasets, allowing more comprehensive studies of the disease. In this work, we introduce a multi-modal lung cancer dataset curated by aggregating data from multiple publicly available biomedical repositories, including CPTAC-LSCC \cite{cptac-lscc}, CPTAC-LUAD \cite{cptac-luad}, TCGA-LUSC \cite{tcga_lusc}, and TCGA-LUAD \cite{tcga_luad}. The resulting cohort comprises 1{,}365 patients, including 1{,}026 from TCGA and 339 from CPTAC, with 1{,}169 patients (86\%) surviving beyond 12 months. Together, these multiple sources form a rich multi-modal resource for studying both Lung Squamous Cell Carcinoma (LSCC) and Lung Adenocarcinoma (LUAD). The dataset integrates multiple data modalities, including clinical records, with demographic and diagnoses related data, transcriptomic records, where a gene selection has been made, and imaging scans including multiple WSI, CT and PET imaging scans, that comprise multiple body parts and multiple scans per person. Moreover, the dataset has longitudinal data with follow-up and treatment information. Modality availability is summarized in Table~\ref{tab:lung-mod}. All patients include clinical data, while WSI  and transcriptomic is available for most and follow-up information for 65\%. In contrast, CT and PET imaging are more sparsely represented, available for 5\% and 2\% of patients, respectively. Additionally, there are four specific treatment modalities: chemotherapy, radiation therapy, surgery and immunotherapy. 68\% of patients had radiation therapy, approximately 20\% had chemotherapy and surgery, and a small cohort of 4\% have had immunotherapy. 

\begin{table}[!htb]
\centering
\caption{Distribution of patients per modality in the lung dataset.}
\label{tab:lung-mod}
\begin{tabular*}{0.8\textwidth}{@{\extracolsep{\fill}}ccccc@{}}
\toprule
{\color[HTML]{000000} Modality} &
  {\color[HTML]{000000} \begin{tabular}[c]{@{}c@{}}Patient\\ Number\end{tabular}} &
  {\color[HTML]{000000} \begin{tabular}[c]{@{}c@{}}Modality\\ Missingness\end{tabular}} &
  {\color[HTML]{000000} \begin{tabular}[c]{@{}c@{}}Alive\\ @12months\end{tabular}} &
  {\color[HTML]{000000} \begin{tabular}[c]{@{}c@{}}Deceased\\ @12months\end{tabular}} \\ \midrule
{\color[HTML]{000000} CLIN}  & {\color[HTML]{000000} 1365} & {\color[HTML]{000000} 0}     & {\color[HTML]{000000} 1169 (86\%)} & {\color[HTML]{000000} 196 (14\%)} \\
{\color[HTML]{000000} TRNSC} & {\color[HTML]{000000} 1284}  & {\color[HTML]{000000} 6\%}  & {\color[HTML]{000000} 1107 (86\%)}  & {\color[HTML]{000000} 177 (14\%)}  \\
{\color[HTML]{000000} FUP}   & {\color[HTML]{000000} 880}  & {\color[HTML]{000000} 35\%}  & {\color[HTML]{000000} 777 (88\%)}  & {\color[HTML]{000000} 103 (12\%)} \\
{\color[HTML]{000000} CHEM}  & {\color[HTML]{000000} 277}  & {\color[HTML]{000000} 80\%}  & {\color[HTML]{000000} 245 (88\%)}  & {\color[HTML]{000000} 32 (12\%)}  \\
{\color[HTML]{000000} RAD}   & {\color[HTML]{000000} 923}  & {\color[HTML]{000000} 32\%}  & {\color[HTML]{000000} 804 (87\%)}  & {\color[HTML]{000000} 119 (13\%)} \\
{\color[HTML]{000000} SURG}  & {\color[HTML]{000000} 291}  & {\color[HTML]{000000} 79\%}  & {\color[HTML]{000000} 250 (86\%)}  & {\color[HTML]{000000} 41 (14\%)}  \\
{\color[HTML]{000000} IMMUNO} & {\color[HTML]{000000} 49}   & {\color[HTML]{000000} 96\%}  & {\color[HTML]{000000} 45 (92\%)}   & {\color[HTML]{000000} 4 (8\%)}    \\
{\color[HTML]{000000} WSI}   & {\color[HTML]{000000} 1359} & {\color[HTML]{000000} 0.4\%} & {\color[HTML]{000000} 1164 (86\%)} & {\color[HTML]{000000} 195 (14\%)} \\
{\color[HTML]{000000} CT}    & {\color[HTML]{000000} 71}   & {\color[HTML]{000000} 95\%}  & {\color[HTML]{000000} 58 (82\%)}   & {\color[HTML]{000000} 13 (18\%)}  \\
{\color[HTML]{000000} PET}   & {\color[HTML]{000000} 33}   & {\color[HTML]{000000} 98\%}  & {\color[HTML]{000000} 26 (79\%)}   & {\color[HTML]{000000} 7 (21\%)}   \\ \bottomrule
\end{tabular*}
\end{table}

\begin{figure}[!htb]
    \centering
    \includegraphics[width=0.8\linewidth]{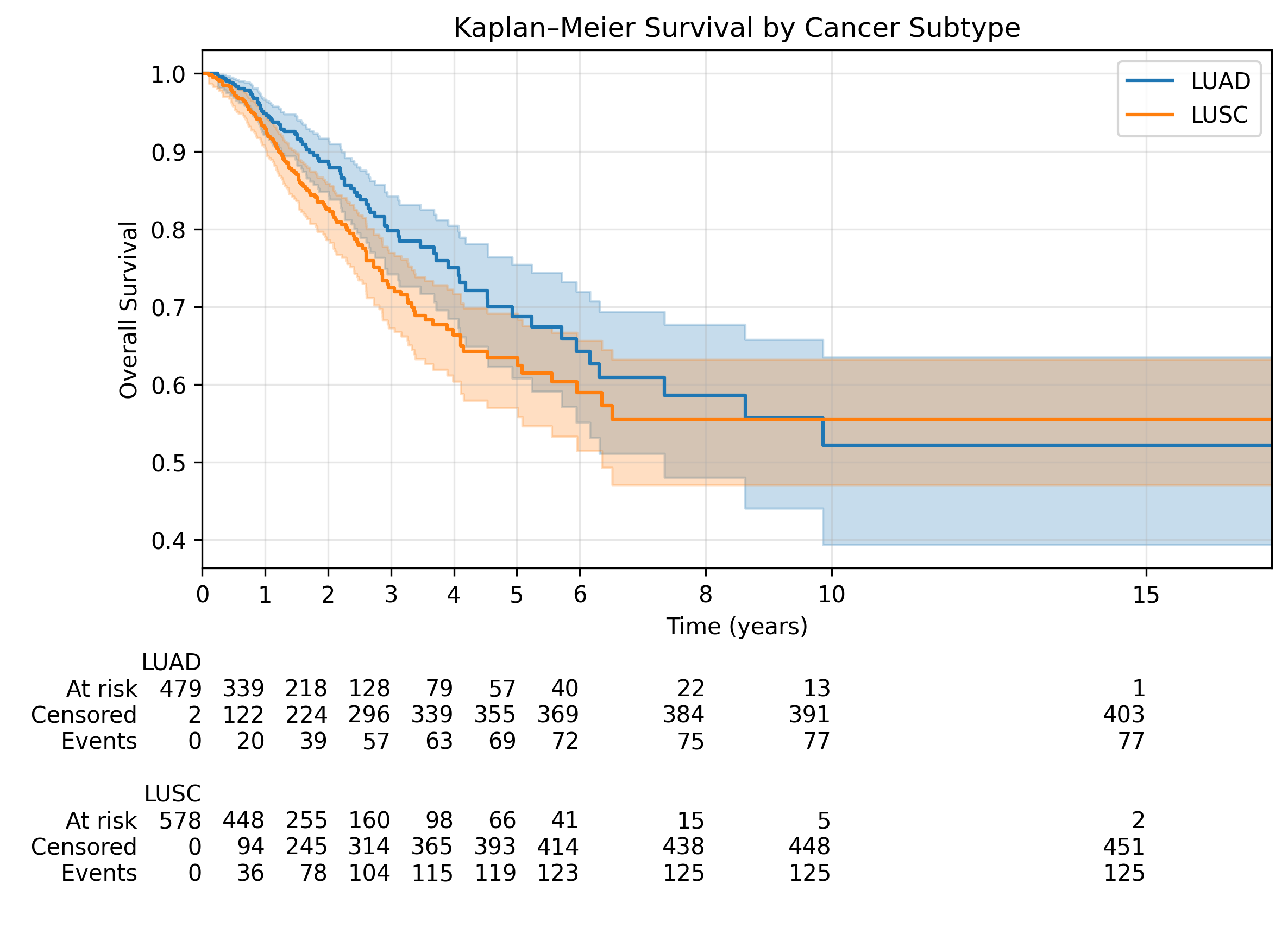}
    \caption{Kaplan-Meier survival plot, stratified by histological subtype.}
    \label{fig:kaplan-meier}
\end{figure}

Furthermore, the dataset provides either the number of days to death or the number of days to the last follow-up visit for each patient. These time-to-event variables enable both survival prediction at user-defined prediction horizons and dynamic survival prediction by incorporating each patient's individual follow-up time. Figure~\ref{fig:kaplan-meier} presents the Kaplan–Meier survival curves stratified by cancer subtype (LUAD and LUSC). This analysis includes 1,059 patients with available survival or follow-up time information. As expected, the number of patients at risk decreases over time due to observed death events (patients who experienced the outcome of interest) and censoring (patients who remained alive or were lost to follow-up before an event was observed), illustrating the longitudinal nature of the cohort. Both histological groups exhibit gradual declines in survival probability over the follow-up period, with substantial numbers of censored observations, indicating that the dataset contains both events and incomplete follow-up. The survival curves also suggest poorer survival for patients with LUSC compared with LUAD. These characteristics make the cohort well suited for survival analysis and dynamic time-to-event prediction, where individual follow-up durations and censoring can be explicitly incorporated into the modeling process.

In the following sections, we provide specific details regarding the curation process of each modality as well as further detail about missingness and their characteristics.

\subsection{Clinical Data}
Clinical data comprise patient demographic characteristics and tumor diagnosis information. Clinical features from the CPTAC and TCGA cohorts were first merged and standardized to resolve inconsistencies in variable names and data types across the two studies. The resulting combined dataset contained a substantial amount of missing data and several redundant variables. Therefore, variables with more than 70\% missing values were excluded, along with patient records lacking vital status information.

The remaining variables were evaluated individually, and those considered most relevant for the analysis were retained, resulting in a final set of 15 clinical features. Of these, 7 correspond to demographic information and 8 to tumor diagnosis. Only 132 patients had complete demographic information, whereas 890 patients had complete diagnosis data, highlighting the high level of missingness in the clinical modality.

Variables were encoded according to their data types. The AJCC tumor staging variables, including primary tumor extent (T), regional lymph node involvement (N), distant metastasis (M), and overall tumor stage (including stages with A and B subcategories), were encoded using ordinal values to preserve the inherent ordering of disease progression. Age was discretized into 10-year intervals. In addition to demographic variables such as ethnicity, race, and gender, the dataset includes smoking-related information, which is particularly relevant for lung cancer.

\subsection{Imaging Data}
The vast majority of patients have WSI data, with a median of 3 slides per patient, resulting in a total of 5{,}427 slides. CT data are available for 71 patients, with a median of 5 scans per patient, totaling 482 scans. PET imaging is the least prevalent modality, available for 33 patients, with a median of 4 scans per patient, yielding 144 total volumes. Even within the same imaging modality, scans may differ in anatomical coverage and acquisition time, making direct alignment or aggregation non-trivial and complicating standardization across samples.
Regarding WSI data, a patient may have multiple whole-slide images corresponding to different specimen types, including diagnostic slides (DX), tissue-side slides (TS), and bottom-side slides (BS). Diagnostic slides are used for the primary histopathological assessment, whereas TS and BS slides are typically acquired during surgical procedures to evaluate whether the resection margins are free of tumor.

The radiology data used in this study (CT and PET) follow a hierarchical structure in which each patient may have multiple imaging studies, each study may contain multiple series, and each series consists of multiple 3D scans in DICOM format. To improve data quality, we excluded non-diagnostic series such as Localizer, Scout, and Reconstructions.

\subsection{Genomics Data}
We collected the genomic data associated with the TCGA and CPTAC patients from the cBioPortal for Cancer Genomics \footnote{https://www.cbioportal.org/}. In detail, the genomic data corresponds to transcriptomics (bulk RNA), which represents raw counts of expression for over 60{,}000 genes. Each patient may be associated with multiple transcriptomic vectors. Thus, to ensure consistency across cases we followed a simple heuristic to select a single vector per patient, based on the sample barcode. The primary tumor sample (designated by the code ”01”) was selected preferentially, as it represents the primary tumor site. For example, given the CPTAC samples ”C3L-02127-01” and ”C3L-02127-06”, the sample ”01” was selected, since ”06” denotes a metastatic site. In instances where the ”01” sample was unavailable, the sample with the lowest numerical identifier was selected, under the assumption that it corresponds to the earliest or most representative available tumor sample. Additionally, samples containing NaN values were considered defective and excluded from the analysis.

In our benchmark experiments, all transcriptomic samples were preprocessed to ensure that the data is comparable across patients. The standard log counts per million normalization was applied, followed by a dimensionality reduction based on the selection of the highly variable genes across the entire cohort, using the methodology described in \cite{stuart2019}, resulting in the selection of a subset of 4{,}096 genes.

\subsection{Follow-Up Data}
The TCGA dataset includes longitudinal follow-up records that capture tumor status information over time, useful to monitor disease progression. In total, 880 patients have follow-up tumor status data, indicating if patient is tumor free or not, comprising 1{,}431 individual records. All of these entries have 'days to followup' variable to enable longitudinal analysis. Each patient has a median of two follow-up visits. The median time to the first follow-up is 298 days, while the second occurs at a median of 647 days.

\subsection{Treatment Data}

The TCGA dataset contains longitudinal treatment records covering chemotherapy, radiation therapy, surgery, and immunotherapy interventions. All selected records include the "days to treatment start" variable, enabling temporal analyses. Some records also provide treatment duration and outcome. Each treatment indicates whether it targets the primary disease, with some further distinguishing synchronous primary tumors, subsequent primaries, recurrence, or metastatic disease. Patients may have multiple treatment records over time.

There are 426 chemotherapy records from 277 patients, of which 209 include treatment outcomes. The first chemotherapy occurs, on average, 155 days after diagnosis, and all records correspond to the primary disease.

The dataset contains 58 immunotherapy records from 49 patients, with outcome information available for 27 records. The first administration occurs on average 440 days after diagnosis, and all treatments target the primary disease.

There are 1{,}337 radiation therapy records from 923 patients, with treatment outcomes reported for 93 records. The first treatment occurs on average 57 days after diagnosis. Of these records, 942 correspond to the primary disease and 395 to non-primary disease.

Finally, there are 525 surgery records from 291 patients. The first procedure occurs on average 599 days after diagnosis. None correspond to the primary disease, and the records specify whether the surgical site is locoregional or distant.

For modeling purposes, treatment variables were transformed into a structured feature representation. A one-hot encoding scheme was used for categorical treatment indicators, while continuous or semi-continuous values were incorporated when dosage information was available.

For immuno and chemotherapy treatments, each drug in the full drug list was represented as a separate feature. When a drug was administered, its dosage was recorded if available (in mg). If administration was confirmed but dosage information was missing, the value was set to -1. If the drug was not administered, the value was set to 0.

Radiation therapy was encoded in a similar manner: the radiation dose was recorded when available (in Gy), and set to -1 when radiation was administered but dose information was missing.

For surgical treatments, information regarding the anatomical surgery sites was split into primary location or distant location.

\section{Benchmarks}
Our curated multi-modal medical dataset supports a wide range of prediction tasks, such as binary survival prediction, which, given the information of the days until death, can be done at any specific horizon. Moreover, the longitudinal nature of the dataset allows to create a more comprehensive overview of the patient, allowing for risk and hazard survival modeling. Furthermore, the dataset enables the investigation of different strategies for handling incomplete multi-modal inputs, such as modality reconstruction or missing-modality masking. In this work, we focus on 12-month overall survival prediction, a widely studied benchmark in medical prognosis, as well as disease-specific survival prediction, both using only data available at diagnosis. In addition, we leverage the longitudinal nature of the dataset by performing another binary disease-specific survival prediction, but using longitudinal modalities to capture the entire patient timeline and predicting survival at a dynamic horizon (last known follow-up appointment for each patient). As well as discrete hazard prediction, modeling the probability of a certain target event occurring across successive time intervals rather than predicting a single binary outcome.

\begin{figure}[t]
    \centering
    \includegraphics[width=0.6\linewidth]{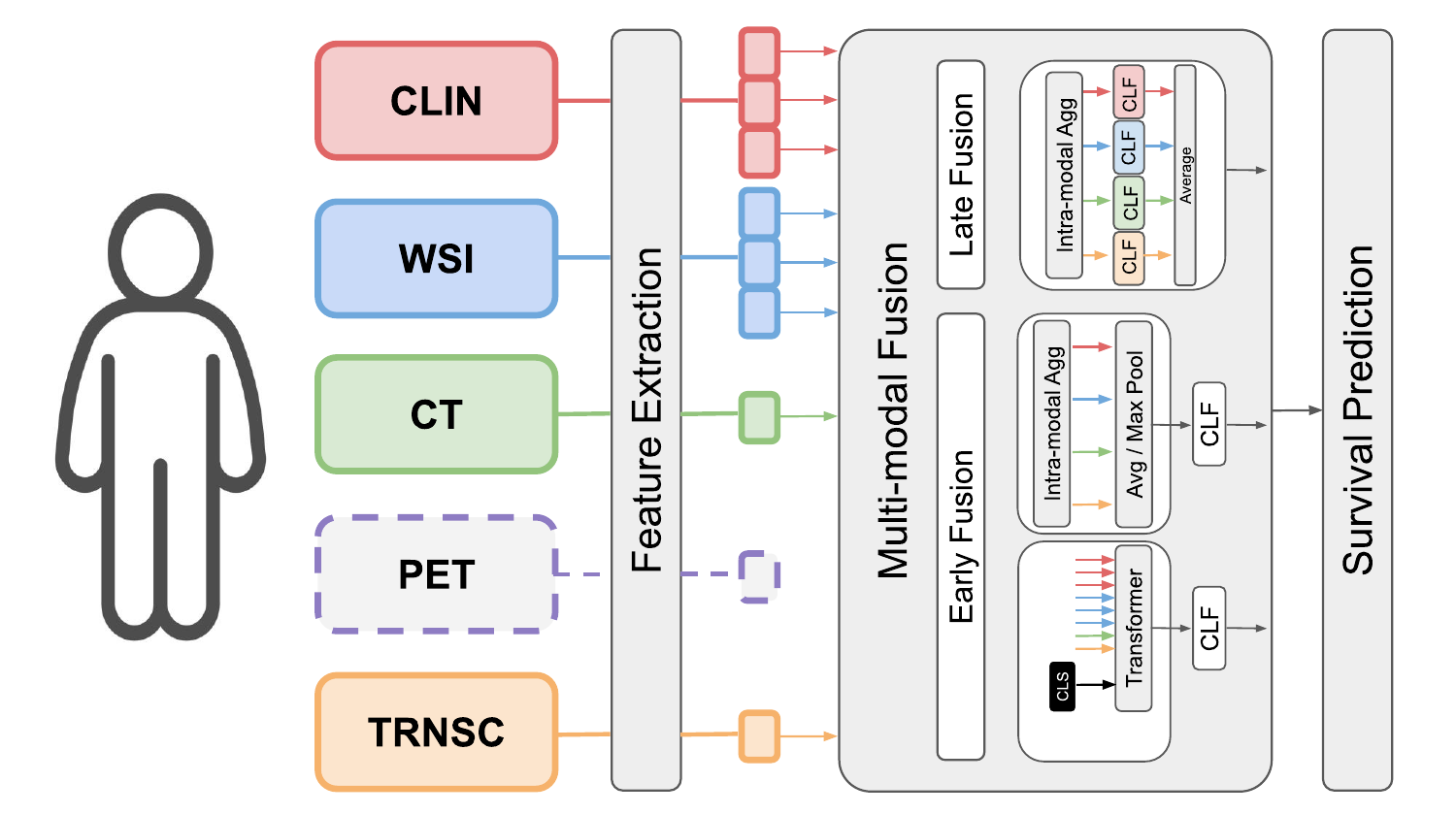}
    \caption{Benchmark pipeline overview. In the example, PET is signaled as missing modality. In our experiments, the classifier (CLF) can be either MLP or XGBoost.}
    \label{fig:transformer-arch}
\end{figure}

We first evaluate the predictive performance of the clinical modality independently, as it is the only modality available for all cohort, followed by multi-modal experiments that assess different fusion strategies as well as different classifiers and the impact of missing data using ablation studies. Specifically, we compare multiple early and late fusion approaches using MLP or XGBoost \cite{xgboost} classifiers. Figure~\ref{fig:transformer-arch} provides an overview of the proposed multi-modal framework. The following sections describe each component in detail.
\subsection{Feature Extraction and Encoding}
Categorical tabular variables are encoded using ordinal encoding whenever an inherent ordering exists, assigning each category an integer that preserves its natural order; otherwise, one-hot encoding is applied. Whole-slide image (WSI) features are extracted using TITAN~\cite{titan}, whereas CT and PET scans are processed with MeD-3D \cite{med3d}, a pre-trained CNN that produces 512 dimensional feature embeddings. The resulting feature maps from the Med-3D were further compressed into a fixed-length feature vector using global 3D average pooling, which aggregates spatial information across the entire volume, ensuring scans of varying sizes are transformed into uniform feature dimensions, suitable for downstream tasks. Following feature extraction, each modality is passed through a dedicated modality-specific encoder that projects its features into a common embedding dimension, enabling subsequent multi-modal fusion.

\subsection{Multi-modal Fusion}
\subsubsection{Early fusion}
We compare three early fusion approaches. Firstly, we employ a Transformer encoder, that fuses modalities by leveraging self-attention. Multiple tokenization strategies can be used to represent each modality (i.e., one token per modality or representing each individual feature as a token). To leverage the most amount of data, we encode imaging modalities as one token per scan. This method also allows for more flexibility when dealing with missing data. By leveraging the Transformer's self-attention masking mechanism, missing modalities can be naturally ignored during attention computation, enabling the model to operate directly on incomplete multi-modal inputs.

We also consider mean and max pooling strategies, in which the sequence of modality embeddings is aggregated using either mean or max pooling to produce a unified patient representation that is passed directly to the classifier. For modalities with multiple instances (e.g., imaging scans), the corresponding embeddings are first averaged to obtain a single modality representation (intra-modal mean instance aggregation). To ensure a fair comparison with the attention-based approach, missing modalities are excluded from the pooling operation rather than contributing to the aggregated representation.

\subsubsection{Late Fusion}
As a late fusion baseline, each modality is processed independently. As in early fusion, multiple-instance modalities are first averaged to obtain a single modality representation. Each representation is then passed through its own classifier to produce a modality-specific prediction. The final patient-level prediction is obtained by averaging the predictions across all available modalities, with missing modalities excluded from the aggregation.

\subsubsection{Classifiers}
We test two different popular classifiers as survival predictors: MLP and XGBoost. Due to high class imbalance, both classifiers are trained using class weights and decision threshold tuning. For MLP, early stopping is used.

\subsection{Longitudinal Modeling}
For the longitudinal tasks, we employ only the Transformer-based fusion model. To capture the longitudinal nature of the data, each modality token is augmented with a learnable time encoding, analogous to the positional encodings used in Transformers. This temporal encoding allows the model to distinguish observations acquired at different time points while preserving their chronological order. Multi-modal fusion is then performed using self-attention, with missing modalities naturally excluded through attention masking. The resulting fused representation is passed to an MLP, whose output layer is adapted to the prediction task: a single binary output for disease-specific survival at dynamic horizon prediction (\ref{sec:4.3}) or a vector of discrete time-bin predictions for survival analysis (\ref{sec:4.4}).  

\subsection{Evaluation Protocol}
The dataset was stratified into five folds at the patient level according to 12-month survival status, such that all CT/PET scans, treatment records, and other data associated with a given patient were assigned to the same fold. All reported results correspond to the average validation performance across the five cross-validation folds. The dataset is highly imbalanced, with approximately 86\% of patients surviving beyond 12 months. Consequently, balanced accuracy (BAcc) is used as the primary evaluation metric for the binary survival prediction task, as it provides a more informative measure of performance under class imbalance. For the longitudinal hazard prediction task, model performance is evaluated using the concordance index (C-index), which measures the model's ability to correctly rank patients according to their risk.

\section{Benchmark Results}
This section presents and discusses the benchmarking results on the lung cancer dataset across four prediction tasks. First, we present results for overall disease-specific survival prediction and 12-month survival prediction. Then, to leverage longitudinal data we predict binary survival prediction at dynamic prediction time horizons, as well as discrete-time hazard prediction. We then compare fusion approaches and interpret benefits of leveraging multi-modal data in this dataset.

\subsection{Disease-Specific Survival Prediction}
\label{sec:4.1}
By incorporating follow-up information on patient tumor status, we evaluate disease-specific survival (DSS), which is often a more clinically meaningful endpoint than overall survival (OS). Unlike OS, DSS considers only deaths attributable to the primary disease, excluding patients who died from unrelated causes. The benchmarking results for this task are presented in Table~\ref{tab:lung-dss}. For this experiment, only information assumed to be available at the time of diagnosis was considered. Consequently, modalities containing explicit longitudinal information were excluded. Therefore, the multi-modal configuration (ALL) comprises clinical, transcriptomic, WSI, CT, and PET data.

By inspecting the results, it is clear that both classifiers exhibit different behaviors in the multi-modal configuration. MLP improves the performance in the multi-modal setting, independently of the modality fusion strategy. On the other hand, XGBoost clearly benefits from a late-fusion framework, where different models are trained, each specialized into one modality.

\begin{table}[t]
\centering
\caption{Benchmarking results for disease-specific survival prediction. CLIN corresponds to uni-modal clinical results. ALL includes CLIN, WSI, Transcriptomics, CT and PET data. \textbf{Bold} highlights best BAcc.}
\label{tab:lung-dss}
\begin{tabular*}{\textwidth}{@{\extracolsep{\fill}}ccccc@{}}
\toprule
 & \multicolumn{2}{c}{\cellcolor[HTML]{FFFFFF}Multi-modal Fusion} & \multicolumn{2}{c}{\cellcolor[HTML]{FFFFFF}Modality Configuration} \\ \midrule
Classifier & Early & Late & CLIN & ALL \\ \midrule

\multirow{4}{*}{\cellcolor[HTML]{FFFFFF}MLP}
& Avg & -   & \multirow{4}{*}{56.74 $\pm$ 4.39} & 69.01 $\pm$ 1.95 \\
& Max & -   &                                      & 69.03 $\pm$ 1.85 \\
& SA  & -   &                                      & 69.93 $\pm$ 2.51 \\
& -   & Avg &                                      & 70.35 $\pm$ 3.12 \\ \midrule

\multirow{3}{*}{\cellcolor[HTML]{FFFFFF}XGB}
& Avg & -   & \multirow{3}{*}{69.07 $\pm$ 0.79} & 61.47 $\pm$ 2.37 \\
& Max & -   &                                             & 60.05 $\pm$ 5.51 \\
& -   & Avg &                                             & \textbf{70.36 $\pm$ 1.82} \\

\bottomrule
\end{tabular*}
\end{table}

An analysis of the importance features obtained with XGBoost and with the Transformer, using attention rollout~\cite{abnar2020quantifying}, highlights two important conclusions. First, several of the most important features are clinically relevant to assess disease severity, such as the American Joint Committee on Cancer (AJCC) TNM system~\cite{amin2017eighth}, which aligns with clinical expectations. Second, the high importance assigned to some of the features, such as patient's race and ethnicity, may result from biases in the dataset. In particular, these biases are associated with features that have severe levels of missingness. While the presence of such biases may be seen as negative, real-world healthcare data is often affected by this issue. Therefore, instead of removing them from the dataset, we purposely kept them in the patient data to promote research in the development of models robust to these biases. 

We also assessed the performance of our benchmark models when one of the clinically meaningful features was fully suppressed in the the test data. The AJCC tumor stage was set to missing and the uni-modal and multi-modal classifiers were applied to the data. The results are shown in Table \ref{tab:ablation}. As expected, removing the AJCC tumor stage generally leads to a decrease in predictive performance across all models. However, the performance gap between the clinical-only and multi-modal (ALL) models becomes more pronounced, indicating that the multi-modal approaches are more robust to the loss of highly informative clinical features by leveraging complementary information from the remaining modalities. This strongly supports the creation of a multi-modal dataset, even when the modalities are not available for all patients. Among the evaluated classifiers, XGBoost exhibits the largest performance degradation, whereas the MLP remains comparatively robust, suggesting a greater capacity to compensate for missing clinical information through multi-modal feature integration.

\begin{table}[t]
\centering
\caption{Results of the study where AJCC tumor stage was set to missing in the disease-specific survival prediction task. \textbf{Bold} highlights best BAcc.}
\label{tab:ablation}
\begin{tabular*}{\textwidth}{@{\extracolsep{\fill}}ccccc@{}}
\toprule
 & \multicolumn{2}{c}{\cellcolor[HTML]{FFFFFF}Multi-modal Fusion} & \multicolumn{2}{c}{\cellcolor[HTML]{FFFFFF}Modality Configuration} \\ \midrule
Classifier & Early & Late & CLIN & ALL \\ \midrule

\multirow{4}{*}{\cellcolor[HTML]{FFFFFF}MLP}
& Avg & -   & \multirow{4}{*}{54.77 $\pm$ 3.15} & 68.39 ± 2.26 \\
& Max & -   &                                      & 67.46 ± 2.09 \\
& SA  & -   &                                      & 64.66 ± 2.11 \\
& -   & Avg &                                      & \textbf{69.93 ± 3.17} \\ \midrule

\multirow{3}{*}{\cellcolor[HTML]{FFFFFF}XGB}
& Avg & -   & \multirow{3}{*}{66.44 ± 1.12}           & 52.24 ± 2.30 \\
& Max & -   &                                             & 49.95 ± 1.78 \\
& -   & Avg &                                             & 55.58 ± 1.79 \\

\bottomrule
\end{tabular*}
\end{table}

\subsection{Survival Prediction at 12 months}
\label{sec:4.2}

Table \ref{tab:lung-results} summarizes the results for 12-month survival prediction using only modalities that do not explicitly incorporate temporal information: transcriptomic, WSI, CT and PET data.

\begin{table}[]
\centering
\caption{Benchmarking results for overall survival prediction at 12 months. ALL includes CLIN, WSI, Transcriptomics, CT and PEt data. \textbf{Bold} highlights best BAcc.}
\label{tab:lung-results}
\begin{tabular*}{\textwidth}{@{\extracolsep{\fill}}ccccc@{}}
\toprule
 &
  \multicolumn{2}{c}{\cellcolor[HTML]{FFFFFF}Multi-modal Fusion} &
  \multicolumn{2}{c}{\cellcolor[HTML]{FFFFFF}Modality Configuration} \\ \midrule
Classifier & Early & Late & CLIN & ALL \\ \midrule

\multirow{4}{*}{\cellcolor[HTML]{FFFFFF}MLP}
& Avg & -   & \multirow{4}{*}{56.67 ± 4.19} & 64.84 $\pm$ 1.81 \\
& Max & -   &                                    & \textbf{65.42 $\pm$ 3.17} \\
& SA  & -   &                                    & 62.93 $\pm$ 1.97 \\
& -   & Avg &                                    & 63.97 $\pm$ 1.80 \\\midrule

\multirow{3}{*}{\cellcolor[HTML]{FFFFFF}XGB}
& Avg & -   & \multirow{3}{*}{64.96 $\pm$ 1.25} & 56.33 $\pm$ 2.89 \\
& Max & -   &                                             & 59.79 $\pm$ 3.53 \\
& -   & Avg &                                             & 59.27 $\pm$ 3.91 \\

\bottomrule
\end{tabular*}
\end{table}

Overall, the results follow a similar trend to the previous task, although with slightly lower predictive performance. While the multi-modal models do not consistently outperform the strongest uni-modal baseline, they again demonstrate greater robustness when highly informative clinical variables are unavailable. This is reported in Table~\ref{tab:ablation-12}, where as before we completely remove the AJCC tumor stage in the inference phase. As expected, performance decreases across all models; however, the multi-modal configuration (ALL) consistently mitigates this degradation, particularly when using the MLP classifier, highlighting the benefit of integrating complementary information across modalities.

\begin{table}[t]
\centering
\caption{Results for the study where AJCC tumor stage was set to missing in the binary survival at 12 months prediction task. \textbf{Bold} highlights best BAcc.}
\label{tab:ablation-12}
\begin{tabular*}{\textwidth}{@{\extracolsep{\fill}}ccccc@{}}
\toprule
 & \multicolumn{2}{c}{\cellcolor[HTML]{FFFFFF}Multimodal Fusion} & \multicolumn{2}{c}{\cellcolor[HTML]{FFFFFF}Modality Configuration} \\ \midrule
Classifier                                    & Early & Late & CLIN                                                   & ALL                   \\ \midrule
\cellcolor[HTML]{FFFFFF}                      & Avg   & -    & \cellcolor[HTML]{FFFFFF}                               & 64.05 ± 1.56          \\
\cellcolor[HTML]{FFFFFF}                      & Max   & -    & \cellcolor[HTML]{FFFFFF}                               & \textbf{64.40 ± 2.90} \\
\cellcolor[HTML]{FFFFFF}                      & SA    & -    & \cellcolor[HTML]{FFFFFF}                               & 61.07 ± 2.58          \\
\multirow{-4}{*}{\cellcolor[HTML]{FFFFFF}MLP} & -     & Avg  & \multirow{-4}{*}{\cellcolor[HTML]{FFFFFF}54.63 ± 4.80} & 62.68 ± 2.43          \\ \midrule
\cellcolor[HTML]{FFFFFF}                      & Avg   & -    & \cellcolor[HTML]{FFFFFF}                               & 48.74 ± 1.43          \\
\cellcolor[HTML]{FFFFFF}                      & Max   & -    & \cellcolor[HTML]{FFFFFF}                               & 48.42 ± 2.18          \\
\multirow{-3}{*}{\cellcolor[HTML]{FFFFFF}XGB} & -     & Avg  & \multirow{-3}{*}{\cellcolor[HTML]{FFFFFF}57.08 ± 2.75} & 51.24 ± 0.86          \\ \bottomrule
\end{tabular*}
\end{table}

\subsection{Survival Prediction at Dynamic Horizon Leveraging Longitudinal Multi-Modal Data}
\label{sec:4.3}
Whereas the previous experiments relied exclusively on information available at diagnosis, this task predicts the patient's disease-specific vital status at the last known follow-up by leveraging the complete longitudinal patient trajectory, including treatment information. A total of 1{,}059 patients were included in this experiment using Clinical, Transcriptomics and all available treatment modalities. Incorporating longitudinal treatment information substantially improved predictive performance over the diagnosis-only setting presented in Sections~\ref{sec:4.1} and \ref{sec:4.2}, achieving a \textbf{BAcc of} $\mathbf{81.73 \pm 0.37\%}$.

To assess whether differences in follow-up duration could explain this improvement, we compared the distribution of days to last follow-up between patients who were alive and those who had died using a Mann-Whitney U test. The test found no statistically significant difference between the two groups ($p = 0.208$). Additionally, a logistic regression model using days to last follow-up as the sole predictor of disease-specific vital status identified only a weak negative association ($p = 0.018$), indicating that longer follow-up was associated with lower odds of disease-specific death. Although such an association is expected in longitudinal clinical datasets, its magnitude is insufficient to account for the substantial improvement in predictive performance, supporting the conclusion that the gain primarily arises from the incorporation of longitudinal patient information.

\subsection{Longitudinal Hazard Survival Analysis}
\label{sec:4.4}
Beyond dynamic outcome prediction, the longitudinal nature of the dataset also enables survival analysis, where the objective is to model the evolution of a patient's risk over time rather than predicting a single endpoint. In this experiment, we formulate the task as a discrete-time hazard prediction problem. The event of interest is disease-specific death, represented by the binary indicator $\delta \in \{0,1\}$.

\begin{figure}[t]
    \centering
    \includegraphics[width=0.7\linewidth]{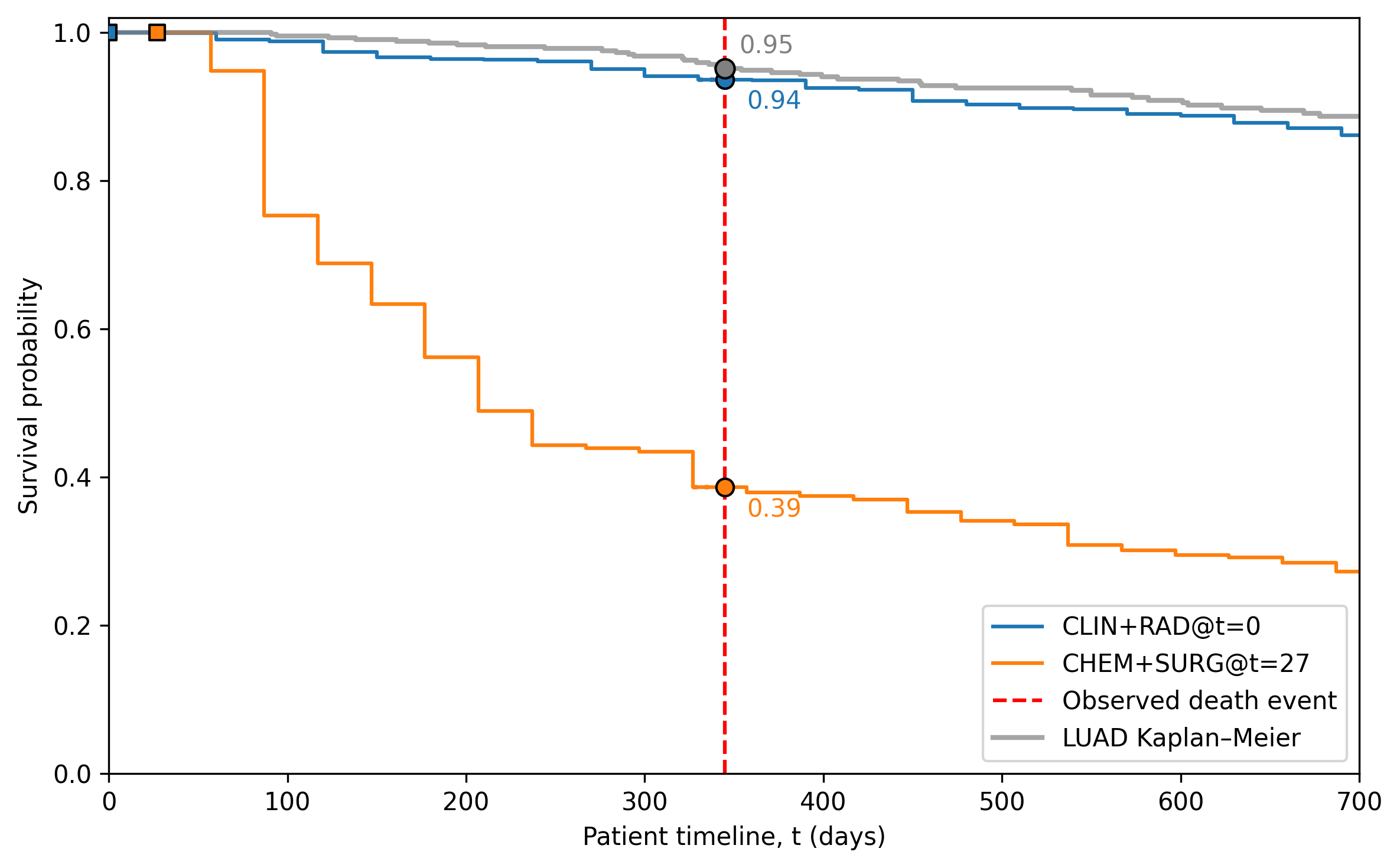}
    \caption{Predicted survival functions for a representative LUAD patient at diagnosis (t = 0) and after incorporation of longitudinal treatment events (t=27 days). The red dashed line denotes the observed event (death) time. Square points indicate new follow-up events in the patient timeline.}
    \label{fig:survival_curves}

\end{figure}

The model estimates the discrete hazard function, corresponding to the conditional probability that the event occurs during a time interval $[t, t+\Delta t]$ given that the patient has remained event-free up to time $t$. Considering 30 day bins until a maximum horizon of 2 years. From the estimated hazard probabilities, the survival function, $S(t)$, can be obtained as the cumulative probability of surviving beyond each time interval, providing an estimate of the patient's survival trajectory over time. The model achieved a \textbf{concordance index of} $\mathbf{85.59 \pm 4.03\%}$. Figure~\ref{fig:survival_curves} illustrates the longitudinal nature of the proposed framework through the survival predictions of a representative patient at two different landmark times. The first prediction is performed at diagnosis ($t=0$), when only baseline clinical and radiological information is available. The second prediction is made after additional longitudinal information, including chemotherapy and immunotherapy treatment events, has been incorporated into the patient's history. As the prediction is conditioned on the updated patient trajectory, the estimated hazard and corresponding survival probabilities are revised, resulting in a substantially lower probability of surviving until the observed event. Notably, the survival curve predicted at diagnosis closely overlaps the Kaplan-Meier survival function, indicating that, based solely on the baseline information, the patient's prognosis is consistent with the average survival pattern observed in the cohort. After incorporating longitudinal treatment information, however, the model revises the prognosis to reflect the patient's evolving clinical trajectory. This example highlights the ability of the proposed framework to continuously update prognostic estimates as new multi-modal longitudinal data become available, rather than relying solely on information available at diagnosis.

\section{Conclusion}
This paper introduces a curated, real-world multi-modal medical dataset comprising over 1{,}300 patients diagnosed with lung adenocarcinoma and lung squamous cell carcinoma. The dataset includes multiple instances of imaging data (WSI, CT, and PET), clinical variables, transcriptomic profiles, as well as longitudinal follow-up and treatment information. Importantly, it exhibits heterogeneous and clinically realistic patterns of missing modalities, closely reflecting real-world practice. We conducted benchmarking on multi-modal binary survival prediction tasks, demonstrating their utility for evaluating multi-modal fusion strategies, methods for handling missing data, and modeling inter-modal interactions. In addition, the temporal structure of the lung dataset enables survival risk modeling over time, capturing tumor trajectories and disease progression dynamics. Overall, our results highlight the value of multi-modal integration for clinical prediction tasks, as well as the utility of this dataset as a challenging benchmark for multi-modal learning under realistic missing-data conditions. We hope this work will facilitate further research in multi-modal fusion, robustness to missing modalities, and temporal clinical modeling, and encourage the community to develop similarly rich datasets.

As an important direction for future work, a benchmark based on nested cross-validation should be established to mitigate the risk of pretraining data leakage and support fairer external comparisons, particularly given the widespread use of TCGA for training medical foundation models.


\section*{Acknowledgements}
This work was supported by LARSyS FCT funding (DOI: \nolinkurl{10.54499/LA/P/0083/2020} and \nolinkurl{10.54499/UID/50009/2025}), and partially funded with grant 2023.02043.BDANA from Fundação para a Ciência e Tecnologia, the FCT project MMIST [2024.06929.RESTART], and the EU project NextGen.

%
%
\bibliographystyle{splncs04}
\bibliography{main}
\end{document}